\documentclass[prl,twocolumn,floatfix,showpacs,preprintnumbers,amsmath,amssymb]{revtex4}

\usepackage[dvips]{graphics}

\usepackage{graphicx}

\begin{document}

\title{Full Counting Statistics of Multiple Andreev Reflections}

\author{J.C.~Cuevas$^1$ and W.~Belzig$^2$}

\affiliation{$^1$Institut f\"ur Theoretische Festk\"orperphysik,
  Universit\"at Karlsruhe, D-76128 Karlsruhe, Germany \\
  $^2$Department of Physics and Astronomy, University of Basel,
  Klingelbergstr.82, CH-4056 Basel, Switzerland}

\date{\today}

\begin{abstract}
  We derive the full distribution of transmitted particles through a
  superconducting point contact of arbitrary transparency under
  voltage bias. The charge transport is dominated by multiple Andreev
  reflections. The counting statistics is a multinomial
  distribution of processes, in which multiple charges $ne$
  ($n=1,2,3,\ldots$) are transferred through the contact. For zero
  temperature we obtain analytical expressions for the probabilities
  of the multiple Andreev reflections. The current, shot noise and
  high current cumulants in a variety of situations can be obtained
  from our result.
\end{abstract}


\maketitle

The complete understanding of the electronic transport in mesoscopic
systems requires information that goes beyond the analysis of the current. 
This explains the great attention devoted in the last years
to current fluctuations in these systems~\cite{Blanter2000}. An
important goal is to obtain the full current distribution. This was
realized by Levitov and coworkers~\cite{Levitov1993}, who borrowed the
concept of full counting statistics (FCS) for photons and adapted it
to electrons in mesoscopic systems. FCS gives the probability $P(N)$
that $N$ charge carriers pass through a conductor in the measuring
time. From the knowledge of these probabilities one can easily derived
not only the conductance and noise, but all the cumulants of the
current distribution. Since the introduction of FCS for electronic
systems, the theory has been sophisticated and applied to many
different contexts (see a recent review, \cite{nazarov:03}).  In
particular one of the authors and Nazarov have shown that, based on a
Keldysh-Green's function method, one can calculate in a unified manner
the FCS of all contacts involving superconducting
elements~\cite{Belzig2001}.

In the context of superconductivity the basic situation, in which the
FCS has not been yet investigated, is a point contact between two
superconductors out of equilibrium. In this system the transport
properties for voltages $V$ below the superconducting gap $\Delta$ are
dominated by coherent multiple Andreev reflections
(MAR)~\cite{Klapwijk1982}. In these processes a quasiparticle
undergoes a cascade of Andreev reflections until it reaches an empty
state in one of the leads. Recently, the microscopic theory of
MAR~\cite{Bratus1995} has provided a new insight into this problem and
has allowed the calculation of properties beyond the current such as
the shot noise~\cite{Cuevas1999}. The predictions of this theory 
have been quantitatively tested in an impressive series of 
experiments in atomic-size
contacts~\cite{Scheer1997,Scheer1998,Cron2001}. In
particular, the analysis of the shot noise~\cite{Cuevas1999,Cron2001}
has suggested, that the current at subgap energies proceeds in
``giant" shots, with an effective charge $q \sim e(1 + 2\Delta/|eV|)$.
However, strictly speaking, the question of whether the charge in
these contacts is indeed transferred in big chunks can only be
rigorously resolved by the analysis of the FCS. This leads us to the
central question addressed in this paper: what is the FCS of MAR?

The answer, which we derive below, is that the statistics is a
\textit{multinomial} distribution of multiple charge transfers.
Technically, we find that the cumulant generating function (CGF) for
a voltage $V$ has the form
\begin{equation}
  \label{eq:marfcs}
  S(\chi) = \frac{t_0}{h}\int_0^{eV} dE \ln \left[
  1+\sum_{n=-\infty}^{\infty}P_n(E,V)\left(e^{in\chi}-1\right)\right]\,.
\end{equation}
The CGF is related to the FCS by $P(N)=\int_{-\pi}^\pi (d\chi/2\pi)
\exp\left[ S(\chi)-iN\chi \right]$. The different terms in the sum in
Eq.~(\ref{eq:marfcs}) correspond to transfers of multiple charge
quanta $ne$ at energy $E$ with the probability $P_n(E,V)$, which can
be seen by the $(2\pi/n)$-periodicity of the accompanying 
$\chi$-dependent counting factor. This is the main result of our work
and it proves, that the charges are indeed transferred in large quanta.

\begin{figure}[t]
\begin{center}
\includegraphics[width=\columnwidth,clip=]{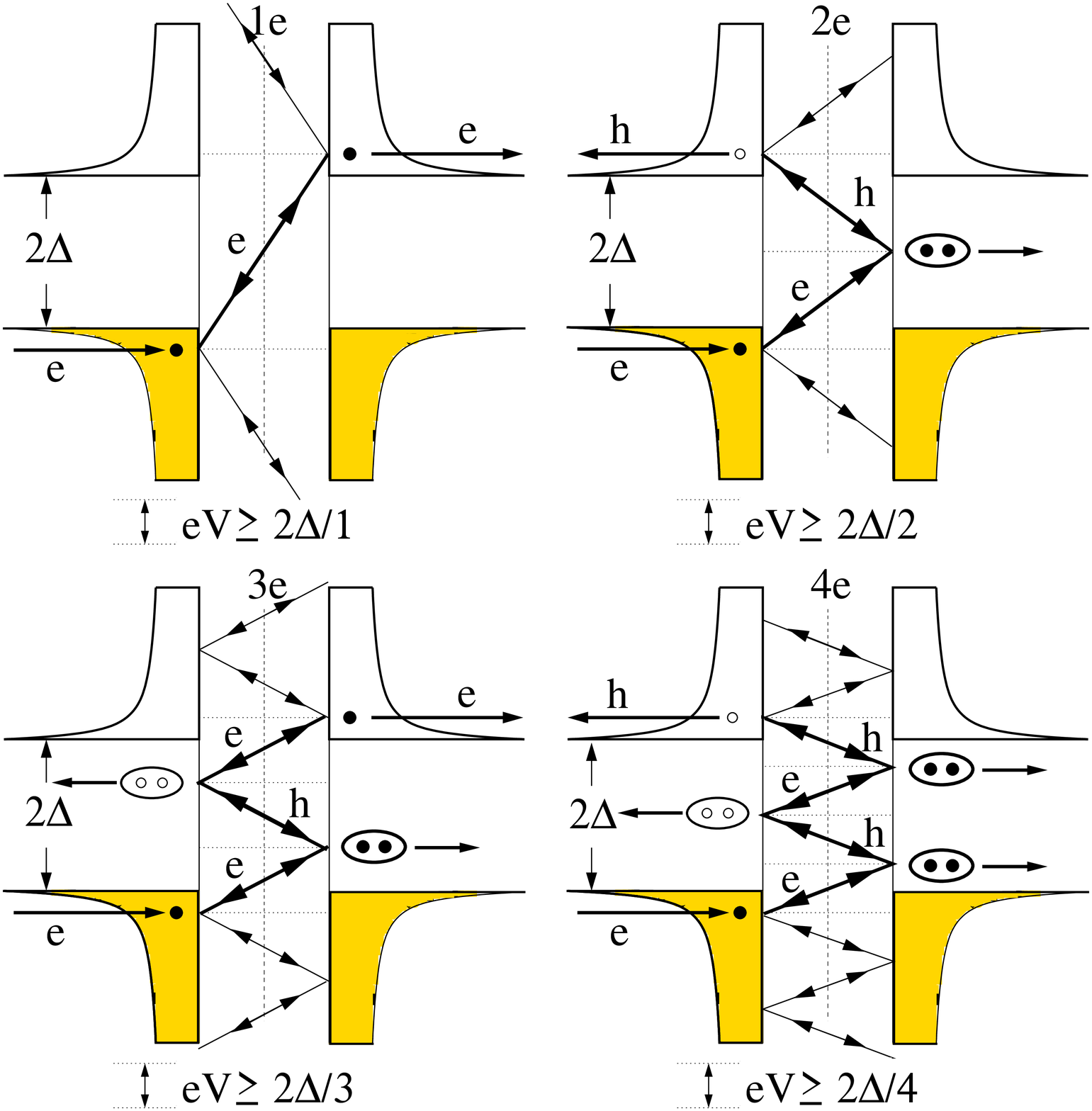}
\caption{Schematic representation of the MARs for BCS
superconductors with gap $\Delta$. We have sketched the density
of states of both electrodes. In the upper left panel we describe
the process in which a single electron tunnels through the system 
overcoming the gap due to a voltage $eV \ge 2\Delta$. The other 
panels show MARs of order $n=2,3,4$. In these processes an incoming
electron at energy $E$ undergoes at least $n-1$ Andreev reflections
to finally reach an empty state at energy $E+neV$. In these MARs
a charge $ne$ is transferred with a probability, which
for low transparencies goes as $T^n$. At zero temperature they
have a threshold voltage $eV=2\Delta/n$. The arrows pointing to the 
left in the energy trajectories indicate that a quasiparticle  
can be normal reflected. The lines at energies below $E$ and above
$E+neV$ indicate that after a detour a quasiparticle can be
backscattered to finally contribute to the MAR of order $n$.}
\label{MAR}
\end{center}
\end{figure}

To arrive at these conclusions, we consider a voltage-biased
superconducting point contact, i.e. two superconducting electrodes
linked by a constriction, which is shorter than the coherence length
and is described by a transmission probability $T$. To obtain the FCS
in our system of interest we make use of the Keldysh-Green's function
approach to FCS introduced Nazarov and one of the
authors~\cite{Nazarov:99,Belzig2001}. The FCS of superconducting
constrictions has the general form \cite{Belzig2001}
\begin{equation}
  \label{eq:cgf}
  S(\chi) = \frac{t_0}{h} {\textrm{Tr}}
  \ln\left[1+\frac{T}{4} 
  \left(\{\check G_1(\chi), \check G_2\}_\otimes -2\right)\right]\;.
\end{equation}
Here $\check G_{1(2)}$ denote matrix Green's functions of the left and
the right contact. The symbol $\otimes$ implies that the products of
the Green's functions are convolutions over the internal energy
arguments, i.e. $(G_1 \otimes G_2) (E,E^{\prime}) = \int dE_1 \;
G_1(E,E_1) G_2(E_1,E^{\prime})$. The trace runs not only over the
Keldysh-Nambu space, but also includes integration energy. For a
superconducting contact at finite bias voltage the CGF depends on time
and Eq.~(\ref{eq:cgf}) is integrated over a long measuring time $t_0$,
much larger than the inverse of the Josepshon frequency.

Let us now describe the Green's functions entering Eq.~(\ref{eq:cgf}).
The counting field $\chi$ is incorporated into the matrix Green's
function of the left electrode as follows
\begin{equation}
  \label{eq:countrot}
  \check G_1(\chi,t,t^{\prime}) = e^{-i\chi\check \tau_K/2}
  \check G_1(t,t^{\prime}) e^{i\chi\check \tau_K/2}\,.
\end{equation}
Here $\check G_1(t,t^{\prime})$ is the reservoir Green's function in the
absence of the counting field and $\check\tau_{\rm
  K}=\hat\sigma_3\bar\tau_3$ a matrix in Keldysh($\hat{\ 
}$)-Nambu($\bar{\ }$) space. We set the chemical potential of the
right electrode to zero and represent the Green's functions by $\check
G_1(t,t^{\prime}) = e^{i eVt \bar \tau_3} \check G_S(t-t^{\prime})
e^{-i eVt^\prime \bar \tau_3}$ and $\check G_2(t,t^{\prime}) =
\check G_S(t-t^{\prime})$. Here, we have not included the dc
part of the phase, since it can be shown that it drops from
the expression of the dc FCS at finite bias. $\check G_S$ is the
Green's function of a superconducting reservoir (we consider the 
case of a symmetric junction), which reads
\begin{equation}
  \label{eq:reservoir}
  \check G_S(E)= \left( \begin{array}[c]{cc}
  (\bar A - \bar R) f + \bar R & (\bar A - \bar R) f \\
  (\bar A - \bar R) (1 - f) & (\bar R - \bar A) f + \bar A
  \end{array}\right).
\end{equation}
Here $\bar R(\bar A)(E)$ are retarded and advanced Green's functions
of the leads and $f(E)$ is the Fermi function. Advanced and retarded
functions in (\ref{eq:reservoir}) have the Nambu-structure $\bar
R(\bar A) = g^{\text{R,A}}\bar\tau_3 + f^{\text{R,A}}\bar\tau_1$
fulfilling the normalization condition $f^2+g^2=1$. They depend on
energy and the superconducting order parameter $\Delta$.

In Eq.~(\ref{eq:cgf}) the matrix appearing inside the logarithm has
an infinite dimension in energy space. In the case of N-N or N-S
contacts such a matrix is diagonal in this space, which makes 
almost trivial the evaluation of the FCS. In the S-S case at finite
bias this is no longer true, which introduces an enormous 
complication. 

We now tackle the problem of how the functional convolution in
Eq.~(\ref{eq:cgf}) can be treated. 
The time-dependence of the Green's functions leads to a representation
of the form $\check G(E,E^{\prime}) = \sum_{n} \check G_{0,n}(E)
\delta(E - E^{\prime} +neV)$, where $n=0,\pm 2$. Restricting the
fundamental energy interval to $E-E^\prime \in [0,eV]$ allows to
represent the convolution as matrix product, i.e.  $(G_1
\otimes G_2) (E,E^{\prime}) \to (\check G_1 \check G_2)_{n,m}
(E,E^{\prime}) = \sum_k (G_1)_{n,k}(E,E^\prime) (G_2)_{k,m}(E,E^{\prime})$.
Writing the CGF as $S(\chi)=(t_0/h) \textrm{Tr} \ln \check Q$, where
$\check Q = 1 + (\sqrt{T}/2) \left( \check G_1(\chi) - 
\check G_2 \right)$~\cite{note}. The trace in 
this new representation is written as $\int_0^{eV} dE \sum_n
\textrm{Tr} \ln \left(\check Q\right)_{nn}$. In this way the
functional convolution is reduced to matrix algebra for the
infinite-dimensional matrix $\check Q$. Still, the task to compute
${\textrm{Tr}} \ln \check Q$ is nontrivial. However, noting that
$\textrm{Tr}\ln\check Q= \ln \det \check Q$, it is obvious at this
stage that $\det \check Q$ has the form of a Fourier series in
$\chi$, which allows us to write the CGF as follows
\begin{equation}
  S(\chi) = \frac{t_0}{h} \int^{eV}_0 dE \; \ln \left[ 
  \sum^{n=\infty}_{n=-\infty} P^{\prime}_n(E,V) e^{in \chi} \right]\,.
\end{equation}
Keeping in mind the normalization $S(0) = 0$, it is clear that one can
rewrite this expression in the form anticipated in Eq.~(1), where the
probabilities are given by $P_n(E,V) = P^{\prime}_n(E,V) /
\sum^{n=\infty}_{n=-\infty} P^{\prime}_n(E,V)$.  Of course, one has
still to extract the expression of these probabilities from the
determinant of $\check Q$, which is a non-trivial task.
It turns out that $\check Q$ has a block-tridiagonal
form, which allows to use a standard recursion technique. We define
the following 4$\times$4 matrices
\begin{eqnarray}
  \check F_{\pm n} & = & \check Q_{\pm n,\pm n} - \check Q_{\pm n,\pm
  n\pm 2} 
  \check F^{-1}_{\pm n \pm 2} \check Q_{\pm n \pm 2,\pm n}  \;;\; n \ge 2 
  \nonumber \\
  \check F_0 & = & \check Q_{0,0} - \check Q_{0,-2} \check F^{-1}_{-2}
  \check Q_{-2,0} - \check Q_{0,2} \check F^{-1}_{2} \check Q_{2,0} ,
\end{eqnarray}
where $\check Q_{n,m}(E) = \check Q(E+neV,E+meV)$. With these
definitions, $\det \check Q$ is simply given by $\det \check Q =
\prod^{\infty}_{j=-\infty} \det \check F_{2j}$. In practice, $\det
\check F_{n} =1$ if $|n| \gg \Delta/|eV|$. This reduces the problem
to the calculation of the determinants of $4 \times 4$ matrices. 

In the zero-temperature limit one can work out this idea analytically
to obtain the following expressions for the probabilities
\begin{widetext}
\begin{eqnarray}
  \label{eq:probabilities}
  P^{\prime}_n(E,V) = 
  \sum^{n-1}_{l=0} K_{-n+l-2,l+2} \times
  J_{-n+l}(E) \times \left[ \prod^{l-1}_{k=-n+l+1} (T/4) |f^A_k|^2 \right] \times
  J_l(E) 
  \;;\;\; n \ge 1 & & \nonumber  \\
  P^{\prime}_0(E,V) = K_{0,0} \times
  \left[Z_{0}^R \left( 1 + \frac{\sqrt{T}}{2} (g^R_0 - g^A_{-1}) - \frac{T}{4}
  (f^A_{-1})^2 B^A_{-2} \right) - \frac{T}{4} (f^R_0)^2 \right]
  \times \Bigg[ R \leftrightarrow A \Bigg] 
\end{eqnarray}
Here, we have used the shorthand $g^{A,R}_n(E) \equiv g^{A,R}(E+neV)$, and defined
\begin{equation}
  Z_{\pm n}^{\alpha}=1 \pm
    \frac{\sqrt{T}}{2} (g^\alpha_{\pm (n+1)} - g^\alpha_{\pm n}) - \frac{T}{4}
    (f^\alpha_{\pm (n+1)})^2 B^\alpha_{\pm (n+2)} \;\;;\;\; n \ge 0 \label{eq:z}\,,
\end{equation}
where $\alpha=R,A$, $K_{n,m} = (\prod_{j=1}^\infty \det \check F_{n-2j})
(\prod_{j=1}^\infty \det \check F_{m+2j})$
and the different functions can be expressed as follows
\begin{eqnarray}
  \label{eq:coefficients}
  \left(B^{\alpha}_{\pm n}\right)^{-1} & = & 1 \pm \frac{\sqrt T}{2}
  (g^\alpha_{\pm n} - g^\alpha_{\pm (n-1)}) - \frac{T}{4} 
  (f^\alpha_{\pm n})^2 / Z^\alpha_{\pm n} \;;\;
  \det \check F_{\pm n} = \prod_{\alpha=A,R} \left[ Z^\alpha_{\pm n}
  ( 1 \pm \frac{\sqrt T}{2} (g^\alpha_{\pm n} - g^\alpha_{\pm (n-1)})
  ) - \frac{T}{4} (f^\alpha_{\pm n})^2 \right] \nonumber \\
  \hspace*{-3cm}
  J_{\pm n} & = & \frac{\sqrt{T}}{2} (g^A_{\pm n} - g^R_{\pm n}) \left[ 
    Z^R_{\pm n} Z^A_{\pm n} - \frac{T}{4} |f^A_{\pm n}|^2 \right] 
  \mp \frac{T}{4} (f^A_{\pm n} - f^R_{\pm n}) \left[ f^R_{\pm n} 
    Z^A_{\pm n} + f^A_{\pm n} Z^{R}_{\pm n} \right] 
\end{eqnarray}
\end{widetext}
Notice that, since at zero temperature the charge only flows in
one direction, only the $P_n$ with $n \ge 0$ survive. 
It is worth stressing that the full information about the 
transport properties of superconducting point contacts is 
encoded in these probabilities. Let us remark
that $P_n(E,V)$ are positive numbers bounded between 0 and 1. 
Although at a first glance they look complicated, they can be 
easily computed and provide the most efficient way to calculate 
the transport properties of these contacts. In practice, to
determine the functions $B^{A,R}_{n}$ and $\det \check F_{n}$, one 
can use the boundary condition $B^{A,R}_{n} = \det \check F_{n} =1$
for $|n| \gg \Delta/|eV|$. For perfect transparency 
$(T=1)$ the previous expressions greatly simplify and the probabilities 
$P_n(E,V)$ can be written as
\begin{equation}
  P_n = \sum^{n-1}_{l=0} (1 - |a_{-n+l}|^2) \left[ \prod^{l-1}_{k=-n+l+1} 
  |a_k|^2 \right] (1 - |a_l|^2) ,
  \label{eq:balistic}
\end{equation}
where $a(E)$ is the Andreev reflection coefficient defined as
$a(E) = -if^R(E) / \left[ 1 + g^R(E) \right]$, and $a_n = a(E+neV)$.

In view of Eqs.~(\ref{eq:probabilities}-\ref{eq:coefficients}) the 
probabilities $P_n$ can be interpreted in the following way. $P_n$ 
is the probability of a MAR of order $n$, where a quasiparticle in 
an occupied state at energy $E$ is transmitted to an empty state at
 energy $E+neV$. The typical structure of the leading contribution
to this probability consists of the product of three terms. First, 
$J_0$ gives the probability to inject the incoming quasiparticle at
energy $E$. The term $\prod^{n-1}_{k=1} (T/4) |f^A_k|^2$ 
describes the cascade of $n-1$ Andreev reflections, in which an
electron is reflected as a hole and vice versa, gaining an energy 
$eV$ in each reflection. Finally, $J_n$ gives the probability to 
inject a quasiparticle in an empty state at energy $E+neV$. In the 
tunnel regime $P_n(E,V) = (T^n/4^{n-1}) \rho_0 \rho_n \prod^{n-1}_{k=1} 
|f^A_k|^2$, $\rho(E)$ being the reservoir density of states.
This interpretation is illustrated in Fig.~\ref{MAR}, where we show
the first four processes for BCS superconductors. The product of 
the determinants in the expression of $P^{\prime}_n$ (see 
Eq.~(\ref{eq:probabilities})) describes the possibility that a 
quasiparticle be reflected and make an excursion to energies below
$E$ or above $E+neV$~\cite{Johansson1999}. In the tunnel regime this
possibility is very unlikely and at perfect transparency is forbidden.
As can be seen in Eq.~(\ref{eq:balistic}), for $T=1$ the quasiparticle 
can only move upwards in energy due to the absence of normal reflection.

\begin{figure}[t]
\begin{center}
\includegraphics[width=\columnwidth,clip=]{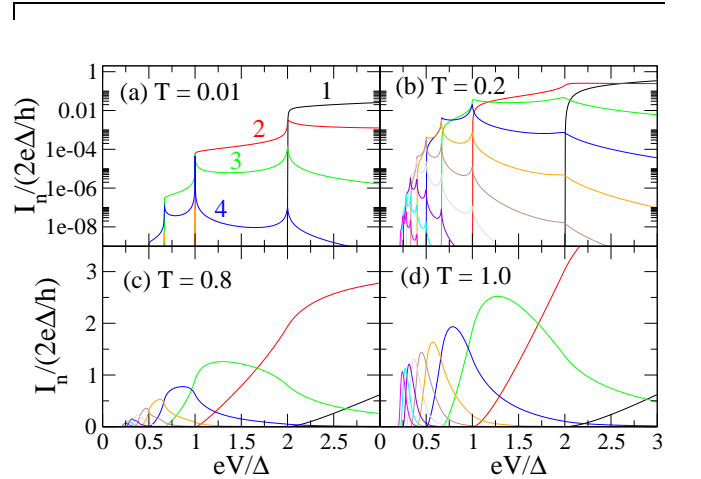}
\caption{Current contribution of processes $n=1,...,10$, from
right to left, as a function of voltage for BCS superconductors of
gap $\Delta$. The different panels correspond to different 
transmissions. Notice the logarithmic scale in the upper panels.}
\label{processes}
\end{center}
\end{figure}

\begin{figure}[t]
  \begin{center}
    \includegraphics[width=\columnwidth,clip=true]{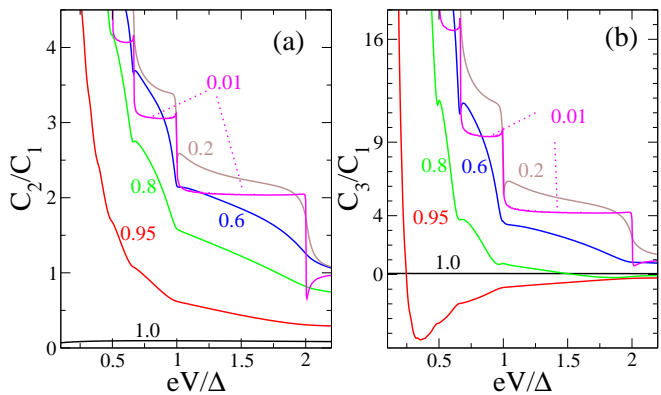}
    \caption{Second and third cumulant at zero temperature for a contact
    between BCS superconductors. Both are normalized to the first
    cumulant (the average current). The transmissions are indicated in
    the plots.} 
\label{cumulants}
\end{center}
\end{figure}

From the knowledge of the FCS one can get a deep insight into the
different transport properties by analyzing the role played
by every process. For instance, in Fig.~\ref{processes} we show 
the contribution to the dc current of the individual processes, 
i.e. $I_n = (2e/h) \int dE\; n P_n$, for the case of BCS 
superconductors of gap $\Delta$. In this case $f^{A,R} = i\Delta 
/ \left[ (E \mp i \delta)^2 - \Delta^2 \right]$, where $\delta = 
0^+$, and $g^{A,R}$ follows from normalization. As can be seen in
Fig.~\ref{processes}, a MAR of order $n$ has a threshold voltage 
$eV=2\Delta/n$, below which it cannot occur. The opening of
MARs at these threshold voltages is the origin of the 
pronounced subgap structure visible in the different transport
properties (see Fig.~\ref{cumulants}). Notice also that at low
transmission the MAR of order $n$ dominates the transport for voltages
$[2\Delta/n,2\Delta/(n-1)]$, while at high transparencies several MARs
give a significant contribution at a given voltage.  This naturally
explains why the effective charge is only quantized in the tunnel 
regime~\cite{Cuevas1999,Cron2001}.

From the CGF one can easily calculate the cumulants of the distribution
and in turn many transport properties. Of special interest are the
first three cumulants $C_1 = \overline N$, $C_2 = 
\overline{(N-\overline{N})^2}$ and $C_3 = \overline{(N-\overline N)^3}$,
which correspond to the average, width and skewness of the distribution,
respectively. From the fact that the FCS is a multinomial distribution, 
it follows that at zero temperature these cumulants can be expressed 
in term of the probabilities $P_n(E,V)$ as $C_n(V) = \int^{eV}_0 dE \;
C_n(E,V)$, where
\begin{eqnarray}
  C_1(E,V) = \sum^{\infty}_{n=1} n P_n \;;\;
  C_2(E,V) = \sum^{\infty}_{n=1} n^2 P_n - C^2_1(E,V) & & 
  \nonumber \\
  C_3(E,V) =  \sum^{\infty}_{n=1} n^3 P_n - C_1(E,V) \left[
  C^2_1(E,V) + 3C_2(E,V) \right] . & & \nonumber 
  \label{eq:cumulants}
\end{eqnarray}
The first two cumulants are simply related to the dc current,
$I = (2e/h) C_1$, and to the zero-frequency noise $S_I =(4e^2/h) C_2$.
In Fig.~\ref{cumulants} we show $C_2$ normalized by $C_1$,
which reproduces the results for the shot noise reported in the
literature~\cite{Cuevas1999}. In this figure we also show $C_3$.
This cumulant determines the shape of the distribution, and it is 
attracting considerable attention~\cite{Levitov2001,Reulet2003}
because it contains information on nonequilibrium physics even at
temperatures larger than the voltage. As seen in Fig.~\ref{cumulants},
at low transmissions $C_3 = q^2 C_1$, where $q(V) = 1 + 
\mbox{Int}(2\Delta/eV)$ is the charge transferred in the MAR which
dominates the transport at a given voltage. This relation is a 
striking example of the general relation conjectured in 
Ref.~\cite{Levitov2001}, and it is simply due to the fact that
the multinomial distribution becomes Poissonian in this limit. 
For higher transmissions this cumulant is negative at high voltage
as in the normal state, where $C_3 = T(1-T)(1-2T)$, but it becomes
positive at low bias. This sign change is due to the reduction of
the MAR probabilities at low voltage. After the sign change there is 
a huge increase of the ratio $C_3/C_1$, which is a signature of the
charge transfer in large quanta. Finally, at $T=1$ the cumulants
($C_n$ with $n > 1$) do not completely vanish due to the fact that 
at a given voltage different MARs give a significant contribution,
and therefore their probability is smaller than one 
(see Fig.~\ref{processes}(d)).

In summary, we have demonstrated that in superconducting
contacts at finite voltage the charge transport is described by
a multinomial distribution of multiple charge transfers. This 
proves that in the MAR processes the charge is indeed transmitted
in large quanta. We have obtained analytically the MAR probabilities
at zero temperature, from which all the transport properties are 
easily computed. Our result constitutes the culmination of the
recent progress in the understanding of MARs, which are a key
concept in mesoscopic superconductivity.

We acknowledge discussions with Yu.V.~Nazarov. JCC was financially
supported by the DFG within the CFN and WB by the Swiss NSF and 
the NCCR Nanoscience.



\begin{thebibliography}{}
\bibitem{note}
  We haved used $2-\{\check G_1,\check G_2\}_\otimes = \left( 
  \check G_1 - \check G_2 \right)^2_\otimes$ to write 
  the CGF as $S(\chi) = (t_0/h) {\textrm{Tr}} \left\{ \ln \check Q_+ 
  + \ln \check Q_- \right\}$, where $\check Q_{\pm} \equiv 1 \pm 
  (\sqrt{T}/2) \left( \check G_1(\chi) - \check G_2 \right)$. 
  Both terms give the same contribution and we concentrate in the 
  analysis of the first one, and we drop the subindex $+$.
\end{thebibliography}
\end{document}